\documentstyle[pra,aps]{revtex}
\begin{document}
\draft \twocolumn[\hsize\textwidth\columnwidth\hsize\csname
@twocolumnfalse\endcsname \title{Decoherence, pointer engineering and quantum
  state protection} 
\author{A.R.R.\ Carvalho, P.\ Milman, R. L. de Matos Filho, and L. Davidovich} 

\address{Instituto de F\'{\i}sica, Universidade Federal do Rio de Janeiro,
  Caixa Postal 68528, 21945-970 Rio de Janeiro, RJ, Brazil}
\date{September 4, 2000}
\maketitle
\begin{abstract}
  We present a proposal for protecting states against decoherence, based on the 
  engineering of pointer states. We apply this procedure to the  vibrational motion of a trapped ion, and show how to protect qubits, squeezed states, approximate phase eigenstates and superpositions of coherent states.
\end{abstract}
 \vskip1pc]

It is well known that the interaction of a quantum system with its surrounding
environment may lead to quantum entanglement between system and environment,
and to an irreversible loss of information on the system.  Which set of states
is less sensitive to entanglement depends on the concrete form of the
interaction Hamiltonian between system and environment \cite{zurek}. In the course of the
interaction, the reduced density operator of the system becomes rapidly
diagonal in this basis, transforming any initial superposition of these states
into a statistical mixture. On the other hand, if the system is initially in
a pointer state, it will remain in a pure state 
during its time development.  The decoherence process by
which coherent superpositions of pointer states get transformed into
statistical mixtures is at the heart of the quantum theory of measurement
\cite{Neumann}, and plays an essential role in the classical limit of quantum
mechanics \cite{classical}. 

Fighting decoherence has become a major chalenge in the last years,
motivated by  recent progress in the theory of quantum
information processing, which relies on the possibility of preserving
quantum coherence \cite{qc,Zoller}. It is also of interest to high-precision frequency measurements in ion traps
\cite{Wineland}.  Several strategies have been devised. They include quantum
error correction schemes \cite{Shor}, feedback implementations
\cite{Mabuchi,Milburn}, the realization of $q$-bits in symmetric subspaces
decoupled from the environment \cite{Zanardi}, and dynamical
decoupling techniques \cite{Viola}.

In linear ion traps, by far the most important decoherence effect is the one
associated with the motional state \cite{Wineland2,James}. In the present
paper, we show that decoherence in the vibrational motion of a trapped ion can
be suppressed by generating, through the techniques of ``reservoir
engineering'' \cite{Zoller2}, artificial reservoirs associated with properly
chosen pointer observables, which have the states to be preserved as their
eigenstates, and which dominate over other dissipation processes.  We exemplify
this procedure by showing how to protect from decoherence several kinds of non-classical states.

Under the hypotheses of Markovian dynamics, complete positivity and initial
decoupling between system and bath \cite{lindblad,alicki}, a master equation
describing the reduced dynamics of a system interacting with its enviroment
can be written in the Lindblad form 
\begin{equation}
  \label{eq:lindblad}
  \frac{d\hat{\rho}}{dt}={\cal L}\hat\rho\equiv\sum_i(\gamma_i/2)\left(2\,\hat{c}_i\,\hat{\rho}\,\hat{c}_i^\dagger -
\hat{c}_i^\dagger\,\hat{c}_i\,\hat{\rho} - \hat{\rho}\,\hat{c}_i^\dagger\,\hat{c}_i\right)\,,
\end{equation}
where $\hat{\rho}$ is the reduced density operator of the system in the
interaction picture, and we have neglected the unitary evolution term $-(i/\hbar)[\hat H,\hat\rho]$. The operators $\hat{c}_i$ are closely related to the
system operators present in the interaction Hamiltonian and $\gamma_i$
measures the strength of the system-enviroment coupling. In this case, the pointer basis is given by the set of all the eigenstates of the
operator $\hat{c}_i$. If all $\hat{c}_i$ are Hermitian, their
eigenstates are steady states of the master equation~(\ref{eq:lindblad}).
If instead the $\hat{c_i}$'s are not Hermitian, the states of the pointer
basis will remain pure, but will not necessarily
be steady states of~(\ref{eq:lindblad}). Note that steady states of
~(\ref{eq:lindblad}) are not affected by the
environment~\cite{Zanardi}. Our strategy for the protection of a
specific quantum state against the  enviroment consists in engineering, 
via adequate external driving of the system,
 a system-enviroment coupling, so that the net effect is to add to the master
equation~(\ref{eq:lindblad}) an extra term, thus getting
\begin{equation}
  \frac{d\hat{\rho}}{dt}={\cal L}\hat\rho+(\Gamma_{\rm eng}/2)\left(2\,\hat{d}\,\hat{\rho}\,\hat{d}^\dagger -
\hat{d}^\dagger\,\hat{d}\,\hat{\rho} - \hat{\rho}\,\hat{d}^\dagger\,\hat{d}\right).
  \label{eq:lindblad2}
\end{equation}
The operator $\hat{d}$ is chosen so that the state one wants to
protect is the {\em only} steady state of 
Eq.~(\ref{eq:lindblad2}) without the environment term ${\cal L}\hat\rho$. For $\Gamma_{\rm eng}\gg\gamma$, the steady
state of the new master equation (\ref{eq:lindblad2}) will be very close to
the state to be protected (if the state is not unique, the term ${\cal L}\hat\rho$ could still induce transitions between the steady states). Besides, {\em any} state of the system will decay into the state chosen to be protected. Therefore, this is also a procedure for preparing quantum states in
the presence of decoherence.

In the following we apply this method to the one-dimensional motion of the center-of-mass
of an ion confined in an electromagnetic trap. The reservoir
engineering process will be implemented by letting the ion interact with
several laser beams of adequate frequencies and intensities, which are quasi-
resonant to an electronic transition $|1\rangle\leftrightarrow|2\rangle$ of
frequency $\omega_{21}$. 

For our purposes it is important to consider the ion to be in the regime of
resolved sidebands, given by $\nu\gg\Gamma, \Omega_n$, where $\Gamma$ is the electronic
energy decay rate, $\nu$ is the ion vibrational frequency, and $\Omega_n$ is the (complex) Rabi frequency corresponding to laser $n$, tuned to the $k$-th
red vibrational sideband of the ion. Under these conditions the interaction term corresponding to laser $n$, can be described, in the interaction picture, by~\cite{vo-ma}
\begin{eqnarray}
  \label{eq:Hint}
  \hat{\tilde{H}}_{\rm int}(t) &=&\hbar g(\hat{A}_{21}\hat d+\hat{A}_{12}{\hat d}^\dagger)\nonumber\\ 
&=&\frac{1}{2}\hbar\Omega_n(i\eta_{n})^k\hat{A}_{21}\hat{f}_k(\hat{\tilde
  a}^\dagger\hat{\tilde a})\,\hat{\tilde a}^k + H.c.,
\end{eqnarray}
where $g$ is taken to be real and
\begin{equation}\label{fk}
  \hat{f}_k(\hat{a}^\dagger\hat{a})=e^{-\eta^2/2}\sum_{l=0}^\infty
\frac{(-1)^l\eta^{2l}}{l!(l+k)!}(\hat{a}^\dagger)^l\hat{a}^l.
\end{equation}
Here, the operators $\hat{a}$ and $\hat{A}_{21}$ are the annihilation operator
of a quantum of the ionic vibrational motion and the electronic flip operator,
respectively. The quantity $\eta_{n}=\sqrt{\hbar ({\bf k}_n\cdot{\bf
    u})^2/2M\nu}$ is the Lamb--Dicke parameter with respect to the direction
of vibration, fixed by the unit vector ${\bf u}$. $M$ is the ion's mass and
${\bf k}_n$ is the wave vector of laser $n$. It is assumed that $\eta\ll1$ 
for any direction orthogonal to ${\bf u}$, as it is the case in
linear traps.

In the Born-Markov limit, the time evolution of the vibronic density operator
$\hat{\rho}$  for the direction $\bf u$ (say $x$) is
\begin{eqnarray}
  \label{eq:Meq}
  \frac{d\hat{\rho}}{dt}&=&-\frac{i}{\hbar}\left[\hat{\tilde{H}}_{\rm int},\hat{\rho}\right]
+\frac{\Gamma}{2}\left(2\hat{A}_{12}\bar{\hat{\rho}}\hat{A}_{21} -
  \hat{A}_{22}\hat{\rho} -\hat{\rho}\hat{A}_{22}\right) \nonumber\\
&+& {\cal{L}}\hat{\rho},
\end{eqnarray}
where the second term corresponds to spontaneous emission with energy
relaxation rate $\Gamma$, and
\begin{equation}
\label{barrho}
  \bar{\hat{\rho}}=\frac{1}{2}\int_{-1}^1\!\!ds\,W(s)e^{i\eta(\hat{\bar{a}}+\hat{\bar{a}}^\dagger)s}\hat{\rho}\,e^{-i\eta(\hat{\bar{a}}+\hat{\bar{a}}^\dagger)s}
\end{equation}
accounts for changes of the vibrational energy along the $x$ direction due to spontaneous emission
with angular distribution $W(s)$.  The last term
of (\ref{eq:Meq}) describes the coupling of the center-of-mass
motion to the environment, and has the general form (\ref{eq:lindblad}).
However, the precise form of this dissipation term is not important for our
purposes.

The matrix elements of (\ref{eq:Meq}) with respect to the electronic basis
yield the equations
\begin{eqnarray}
\dot{\hat{\rho}}_{11}&=&-ig(\hat{d}^\dagger\hat{\rho}_{21}-\hat{\rho}_{12}\hat d)+\Gamma\bar{\hat\rho}_{22}+{\cal L}\hat{\rho}_{11}\,,\label{rho11}\\
\dot{\hat{\rho}}_{22}&=&-ig(\hat d\hat{\rho}_{12}-\hat{\rho}_{21}\hat{d}^\dagger)-\Gamma\hat{\rho}_{22}+{\cal L}\hat{\rho}_{22}\,,\label{rho22}\\
\dot{\hat{\rho}}_{12}&=&-ig(\hat{d}^\dagger\hat{\rho}_{22}-\hat{\rho}_{11}\hat{d}^\dagger)-{\Gamma\over2}\hat{\rho}_{12}+{\cal L}\hat{\rho}_{12}\,,\label{rho12}
\end{eqnarray}

We assume now that the decay rate $\Gamma$ is by far the largest rate in the
system. Under
this condition, one can eliminate $\hat{\rho}_{12}$ adiabatically. Since ${\cal L}\propto\gamma$, one gets:
\begin{equation}\label{rho12s}
\hat\rho_{12}=-(2ig/\Gamma)(\hat{d}^\dagger\hat{\rho}_{22}-\hat{\rho}_{11}\hat{d}^\dagger)\left[1+O\left(\gamma/\Gamma\right)\right]\,.
\end{equation}
Replacing (\ref{rho12s}) into (\ref{rho11}) and (\ref{rho22}), and adding up
these two equations, we get, since the reduced density operator for the
vibrational mode is given by $\hat\rho_v=\hat\rho_{11}+\hat\rho_{22}$, and
neglecting the correction proportional to $\gamma/\Gamma$ in (\ref{rho12s}):
\begin{eqnarray}\label{rhov}
&&\dot{\hat{\rho}}_v={2g^2\over\Gamma}\Big[\left(2\hat d\hat\rho_{11}\hat d^\dagger-\hat d^\dagger\hat d\hat\rho_{11}-\hat\rho_{11}\hat d^\dagger\hat d\right)\\
&&+\left(2\hat d^\dagger\hat{\rho}_{22}\hat d-\hat d\hat d^\dagger\hat{\rho}_{22}-\hat\rho_{22}\hat d\hat{d}^\dagger\right)\Big]
-\Gamma(\hat\rho_{22}-\hat{\bar\rho}_{22})+{\cal L}\hat\rho_{v}\,.\nonumber
\end{eqnarray}

Under the conditions assumed here, the matrix
elements of $\hat\rho_{22}$ are much smaller than those of $\hat\rho_{11}$.
Indeed, replacing (\ref{rho12s}) in (\ref{rho22}), and eliminating
$\hat\rho_{22}$ from (\ref{rho22}) adiabatically, one gets
$\hat\rho_{22}\approx O\left[(g/\Gamma)^2\right]\hat\rho_{11}$. We can
therefore safely neglect the terms dependent of $\hat\rho_{22}$ inside the
brackets in (\ref{rhov}) and at the same time replace $\hat\rho_{11}$ by
$\hat\rho_v$. We verified numerically that these are indeed
excellent approximations. We get then, finally:
\begin{eqnarray}\label{mev}
\dot{\hat\rho}_v&=&{2g^2\over\Gamma}\left(2\hat d\hat\rho_v\hat d^\dagger-\hat d^\dagger\hat d\hat\rho_v-\hat\rho_v\hat d^\dagger\hat d\right)-\Gamma\left(\hat\rho_{22}-\bar{\hat\rho}_{22}\right)\nonumber\\
&+&{\cal L}\hat\rho_v\,.
\end{eqnarray}

We will base our considerations on this equation. The first term on the r.h.s. has the form (\ref{eq:lindblad}). This is the ``engineered reservoir,'' with a decay constant  $\Gamma_{\rm eng}=4g^2/\Gamma$. 

Neglecting terms of O($\eta^4$) in the expansion of
the second term on the r.h.s. of eq.~(\ref{mev}), one can show that its contribution is 
$\sim (\eta^2/5)(4g^2/\Gamma)\hat\rho_{v}$, that is, $(2\eta^2/5)$ multiplied by the engineered-reservoir term. For $\eta\approx
0.25$, this yields a factor $\sim1/40$, a small
correction, which is however fully taken into account in our numerical
simulations.  Therefore, the action of the engineered reservoir will be the
dominant one as long as $\Gamma_{\rm eng}\gg \gamma$.

In recent experiments with trapped ions, random fields seem to play an
important role in the decoherence process \cite{James}. Their effect may also
be described by (\ref{eq:lindblad}). We write the random
field as $E(t)={\cal E}^{(+)}(t) \exp(-i\nu t)+{\cal E}^{(-)}(t)\exp(i\nu t)$,
and the interaction Hamiltonian in the RWA as \cite{James}: $\hat
H=-\mu\left[{\cal E}^{(+)}(t)\hat a^\dagger+{\cal E}^{(-)}(t)\hat a\right]$.
Iterating the equation of motion for the density operator, and using that for
steady fields \cite{Mandel} $\langle{\cal E}^{(-)}(t){\cal
  E}^{(-)}(t')\rangle=\langle {\cal E}^{(+)}(t){\cal E}^{(+)}(t')\rangle=0$,
we get, in the Markovian limit $\langle{\cal E}^{(+)}(t) {\cal
  E}^{(-)}(t')\rangle=2D\delta(t-t')$:
\begin{eqnarray}\label{random}
\dot{\hat{\rho}}&=&(\mu^2D/\hbar^2)\big(2\hat a\hat\rho\hat a^\dagger
-\hat a^\dagger\hat a\hat\rho - \hat\rho\hat a^\dagger \hat a \nonumber \\
&+& 2\hat a^\dagger\hat\rho\hat a - \hat a \hat a^\dagger\hat \rho -\hat
\rho \hat a \hat a^\dagger\big)\,,
\end{eqnarray}
which corresponds to an infinite temperature thermal reservoir (by letting the thermal photon number $N_T\rightarrow\infty$, and at the same time the dissipation rate $\gamma\rightarrow0$, so that $\gamma N_T$ remains constant). Both random fields and thermal reservoirs will be considered in our simulations.

In order to protect a state $|\psi\rangle$, we look for $\hat d$ such that $\hat d|\psi\rangle=\lambda|\psi\rangle$ with $\lambda=0$, and  make sure that $|\psi \rangle$ is in fact the only steady state of
(\ref{eq:lindblad2}) without ${\cal L}\hat\rho$. Since $\Gamma_{\rm eng}\gg\gamma$, this yields a good approximation of the corresponding steady state of (\ref{eq:lindblad2}).

As a first example, we consider the protection (and generation) of the class
of states $|\psi\rangle=\sum_{n=0}^N c_n|n\rangle$, where $|n\rangle$ is an
energy eigenstate of the vibrational motion of the trapped ion and $c_n\neq0$.
It is easy to see that the operator $\hat d=\hat{\rm g}(\hat{n})\,\hat a +
\hat{\rm h}(\hat n)$ has $|\psi\rangle$ as its only eigenstate with eigenvalue
$\lambda=0$, provided the eingenvalues of $\hat{\rm g}(\hat{n})$ and $\hat{\rm h}(\hat{n})$ fulfill the constraints ${\rm g}(m)=-\frac{{\rm
    h}(m)}{\sqrt{m+1}}c_m/c_{m+1}$ ($m=0,\cdots,N\!-\!1$) and $N$ is the first
zero of ${\rm h}(m)$. For this case, $|\psi\rangle$ is the only steady state of (\ref{eq:lindblad2}).

Inspection of
Eqs.~(\ref{eq:Hint}) and (\ref{fk}) shows that the operator
$\hat{\rm g}(\hat{n})\,\hat{a}$ can be engineered by driving the ion with $N$ laser
fields, tuned to the first vibrational sideband of the ion. The values of the
Rabi frequencies $\Omega_n$ of the $N$ lasers are given by the following
system of N linear equations ($m=0,\cdots,N\!-\!1$): 
\begin{equation}
\label{Rabis}
\sum_{n=1}^Ne^{-\eta^2_n/2}\eta_n\Omega_n\sum_{l=0}^m\frac{(-1)^l\eta^{2l}}{l!(l\!+\!1)!}\frac{m!}{(m\!-\!l)!}=\frac{i\rm{h}(m)}{\sqrt{m\!+\!1}}\frac{c_m}{c_{m+1}},
\end{equation}
where the Lamb--Dicke parameters $\eta_n$ depend on the orientation of the laser fields with respect to the
$x$-direction.  The operator $\hat{\rm h}(\hat{n})$ is constructed by driving the ion with two laser fields resonant with the
electronic transition, one of them propagating orthogonal to the $x$-axis
(say $y$-direction, with $\eta_y\ll1$). The Rabi frequencies of these two lasers are related by $\Omega_y=-\Omega_xL_N(\eta_x^2)$, where $L_N(x)$ is a Laguerre
polynomial of order $N$ ($\eta_x$ should not be too large for the first zero of $h(m)$ to occur at $m=N$).

An important representative of the class of states presented above is the ``qubit" state $|\psi\rangle=c_0|0\rangle+c_1|1\rangle$. The
discussion above implies that this state can be generated and protected against the
action of an external reservoir with just three lasers, with Rabi frequencies satisfying the following condition:
\begin{equation}\label{conditions}
{\eta\Omega_x \over i \Omega_1 } = - {c_1 \over c_0 } \quad {\rm and}\quad {\Omega_y \over i \Omega_1 }= e^{-\eta^2/2}{c_1 \over c_0}{{1-\eta^2} \over \eta}.
\end{equation}
In this case, $\Gamma_{\rm eng}=\eta^2\Omega_1^2/\Gamma$ in
Eq.~(\ref{eq:lindblad2}).  In order for the corresponding reservoir to win over
the environment reservoir, one needs $\eta^2{\Omega_1}^2/\Gamma\gg\gamma$, but
at the same time $\Gamma,\Omega_1\ll\nu$ and $\Gamma\gg\eta\Omega_1$. This
requirements are satisfied if $\Gamma\approx4$ MHz, $\Omega_1\approx2$ MHz,
$\eta=0.2$, $\nu\approx20-30$ MHz, as long as $\gamma\ll \eta^2{\Omega_1}^2/\Gamma\approx 40$
kHz. Fig.\ \ref{fig1}a displays the fidelity ${\rm
  F}(t)={\rm Tr}\{\hat{\rho}(0)\hat{\rho}(t)\}$,  with the ion initially in the vibrational state
$|\psi\rangle=\frac{1}{2}(|0\rangle+|1\rangle)$ [all our numerical simulations are obtained from eq.~(\ref{eq:Meq}), and we always assume the ion to be initially in the electronic ground state]. Both  a thermal
and a random field reservoir have been
considered.  As can be seen, the system rapidly reaches a steady state
with fidelity very close to unity ($\eta=0.2$).
 
One should remark that feedback procedures \cite{Milburn} do not protect
states involving superpositions of $|0\rangle$ and $|1\rangle$, since the loss
of one photon by the state $|1\rangle$ completely erases any phase information
about the original state. Our procedure works however very well in this case.

Other interesting example is the
approximate phase state \cite{Pegg}
$|\psi\rangle=(1/\sqrt{N+1})\sum_{n=0}^Ne^{in\phi}|n\rangle$, which can
be generated and protected by $N+2$ lasers.  Fig.\ \ref{fig1}b
displays the time evolution of the fidelity ${\rm F}(t)$,  for the
approximate phase state
$|\psi\rangle=(1/2)\sum_{n=0}^3e^{in\phi}|n\rangle$ ($\eta$'s in the range $0.1-0.2$).

A class of states which is specially fragile against the action of decoherence
is the one formed by mesoscopic superpositions of coherent states. Under
action of an external reservoir, these states decay to a mixture of coherent
states in a extremely short time, inversely proportional to the distance between the two states in phase space \cite{classical}. Our technique can also be applied to the Schr\"odinger-cat-like state \cite{YS} $|\phi_+\rangle=(|\alpha\rangle+i|-\alpha\rangle)/\sqrt{2}$.
Since $|\phi_+\rangle$ has no ``hole" in its number
distribution (which coincides with that for the coherent state $|\alpha\rangle$), it can be approximated by one of the states $|\psi\rangle$ discussed above. Consequently, one could, with the
use of $N+2$ lasers, generate and protect the state 
$|\psi\rangle=\sum_{n=0}^N c_n|n\rangle$, with the first $N$ coefficients $c_n$
equal to the corresponding coefficients of $|\phi_+\rangle$. One should notice, however, that it is possible in this case to find directly a Lindblad operator $\hat{d}$, which has the state $|\phi_+\rangle$
as its only eigenstate with zero eigenvalue: $\hat d= {\cal T}\hat a{\cal T}^\dagger=e^{i\pi \hat
  n} \hat a + i\alpha$. Here ${\cal T}$ is the unitary operator $\exp[i\pi\hat n(\hat n-1)/2]\exp(\alpha \hat a^\dagger-\alpha^* \hat a)$,  which yields $|\phi_+\rangle$ when applied to the vacuum. For this choice of $\hat d$, we plot in
Fig.~1c the fidelity $F(t)$, for  the initial state $|\phi_+\rangle$ ($\eta=0.2$). An open problem is how to engineer this operator with a finite number of laser beams. 

Finally, we describe the protection of a squeezed state. We set $\hat d= \hat a+\chi\hat a^\dagger$, where $\chi=\tanh r$ and $r$ is the squeezing factor. The corresponding setup consists of two lasers along the direction of squeezing, resonant with the first red (laser 1) 
and the first blue (laser 2) sidebands, and with Rabi frequencies satisfying  $\Omega_2/\Omega_1=\chi$ \cite{Zoller2}. The numerical simulation is shown for $r=0.6$ in Fig.\ 1d, for a realistic set of parameters ($\eta=0.05$). Higher values of squeezing render our method less effective, since the presence of higher photon numbers would lead to stronger dissipation by the ``natural'' reservoir.

In conclusion, we have suggested a method for protecting quantum states of the vibrational motion of a trapped ion against
decoherence by generating artificial reservoirs which have the states to be
protected as pointer states. More general pointer states can be generated by applying unitary transformations to the states and operators  discussed here. Indeed, the transformed states would still be the sole steady solutions of the master equation with the transformed operators.  This is precisely the mechanism which leads to the protection of the states $|\phi_+\rangle$ above, and also of squeezed states, since they are related by unitary transformations to the vacuum, which is the only steady state for a zero-temperature reservoir (for which $\hat d=\hat a$). As for possible sources of error in experimental implementations of our proposal, one should remark that our method is sensitive only to relative intensity and phase fluctuations, since state selection is determined by ratios of Rabi frequencies.

We acknowledge the partial support of Conselho Nacional
de Desenvolvimento Cient\'\i fico e Tecnol\'ogico (CNPq), Funda\c c\~ao de
Amparo \`a Pesquisa do Estado do Rio de Janeiro (FAPERJ), Funda\c c\~ao
Universit\'aria Jos\'e Bonif\'acio (FUJB), and Programa de Apoio a
N\'ucleos de Excel\^encia (PRONEX).

\begin{figure}
\caption{Time evolution of the fidelity $F(t)$ for several initial nonclassical states, in the presence of a thermal 
  $\alpha^2=3$; and (d) vacuum squezed state with $r=0.6$.}
\label{fig1}
\end{figure}

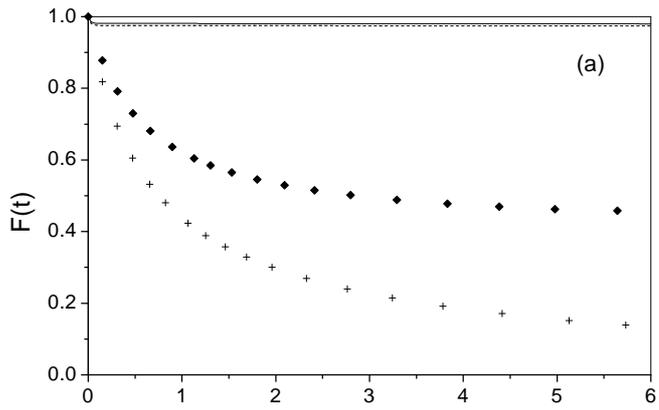
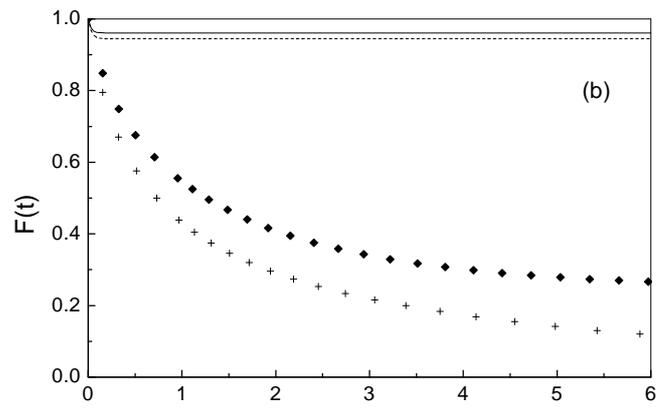
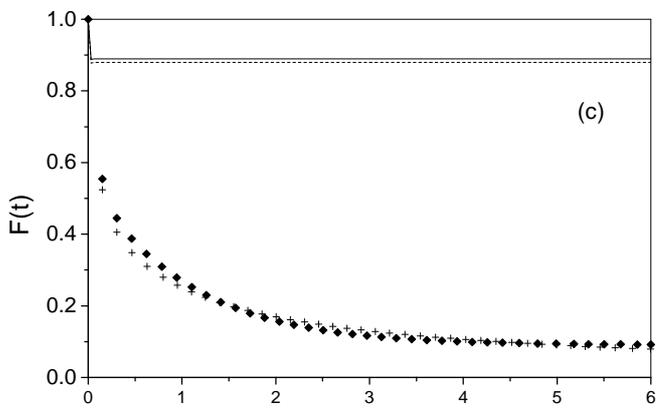
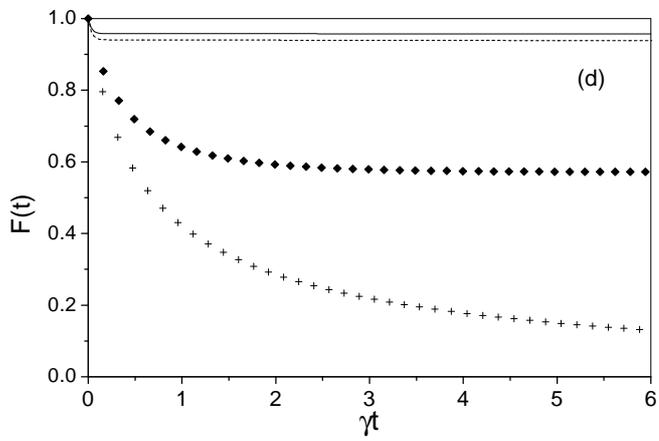
\end{document}